\documentclass[11pt,twoside]{article}
\usepackage{amssymb}
\usepackage[mathscr]{eucal}
\usepackage{eufrak}
\usepackage{amsmath,amsthm}
\usepackage{mathrsfs}
\usepackage[colorlinks=true,
   urlcolor=blue,           filecolor=green,      
   citecolor=green,      
   linkcolor=red,           bookmarks=true,
  unicode,
   plainpages=false,   ]{hyperref}

\usepackage{color}

\def\bp{\begin{proof}}
\def\ep{\end{proof}}
\def\n{\nabla}
\def\ssfrac#1#2{\mbox{\large$\frac{#1}{#2}$}}
\def\sfrac#1#2{\mbox{\Large$\frac{#1}{#2}$}}
\def\intl#1{\int\limits_{#1}}
\def\intll#1#2{\int\limits_{#1}^{#2}}
\def\dm{|\hskip-0.05cm|}

\def\OO{\Omega}
\def\displ{\displaystyle}
\def\VSE{\vspace{6pt}\\&\displ }
\def\VS{\vspace{6pt}\\\displ }
\def\rf#1{{\rm(\ref{#1})}}
\def\R{\Bbb R}
\def\N{\Bbb N}

\def\be{\begin{equation}}
\def\ba{\begin{array}}
\def\ea{\end{array}}
\def\ee{\end{equation}}
\def\ov{\overline}
\def\po{{\partial\Omega}}

\newtheorem{lemma}
{\bf Lemma} 
\font\sc=cmcsc10
\begin{document}
\newtheorem{lem}{\bf Lemma}[section]
\newtheorem{ass}
{\bf Assumption}[section] 
\newtheorem{defi}
{\bf Definition}[section] 
\newtheorem{tho}
{\bf Theorem}[section]  
\newtheorem{rem}
{\sc Remark}[section]  
\newtheorem{coro}
{\bf Corollary}[section]  
\newtheorem{prop}
{\bf Proposition}[section] 
\renewcommand{\theequation}{\arabic{section}.\arabic{equation}}
\title{\Large\bf On Leray's  {\it\textbf {Th\'{e}or\`{e}me de Structure}} for a weak solution\\to the Navier-Stokes equations:\\an improvement of the result} 
\medskip\bigskip
\author{Paolo Maremonti\thanks{Dipartimento di Matematica e Fisica,  
Universit\`{a} degli 
Studi della Campania
``L. Vanvitelli'', via Vivaldi 43, 81100 \null\hskip0.55cmCaserta,
 Italy.\newline\null\hskip0.55cm
paolo.maremonti@unicampania.it \newline\null\hskip0.55cm The  research activity  is performed under the
auspices of   GNFM-INdAM. }}
\date{\it to Franco Oliveri on his 60th birthday}
\markboth{\footnotesize\rm    P. Maremonti} {\footnotesize\rm On\,Leray's\,Th\'{e}or\`{e}me\,de\,Structure\,for\,a\,weak\,solution\,to\,the\,Navier-Stokes\,equations:\,an\,improvement\,of\,the\,result
} 
\maketitle
{\small\bf Abstract} -{ \small In this note we furnish a new result concerning the well known {\it th\'{e}or\`{e}me de structure} by Leray related to a weak solution to the Navier-Stokes equations. We are able to furnish a new estimate on the left endpoint of the unbounded interval of regularity.}
\section{\large Introduction}  We consider the initial boundary value problem for the Navier-Stokes equations:
\be\label{NSD}\ba{c} v_t+v\cdot\n v+\n\pi=\nu\Delta v\,,\;\n \cdot v=0\,,\mbox{ in }(0,T)\times\OO\,,\VS v=0\,,\mbox{ on }(0,T)\times\po\,,\;v=v_0\mbox{ on }\{0\}\times\OO\,.\ea\ee
In \rf{NSD} $\OO\subset\R^3$ is a smooth bounded domain, we set $v_t:=\frac{\partial}{\partial t}v$\,, $v\cdot\n v:=(v\cdot\n) v$\,, and $\nu$ denotes the kinetic viscosity  of the fluid.\par In order to make consistent some quantity that we introduce in the following, we are going to consider problem \rf{NSD} in dimensionless form. Hence, setting $$\texttt{L}:=\texttt{diam}\hskip0.05cm(\OO)\,, \; \texttt{V}:=\Big[{|\OO|}^{-1}\intl\OO |v_0(x)|^2dx\Big]^\frac12\,, \mbox{ and }\texttt{T}:=\ssfrac {\texttt{L}}{\texttt{V}}\,,$$ we consider $\texttt L$, $\texttt V$  and $\texttt T$ as  characteristic length, velocity and time, respectively. We define the Reynolds number $R=\frac{\texttt L\texttt V}\nu$ and the velocity $v':=\frac v{\texttt V}$. With abuse of notation,  we use the same symbols $t,\,x,$ $v$, $(0,T)$ and $\OO$ in place of the dimensionless quantity $x':=\frac x{\texttt L}$, $t':=\frac t{\texttt T}$, $v':=\frac v{\texttt V}$, with $(0,T')$ and $\OO'$ homothetic image domains. Hence, we study the following initial boundary value problem:
\be\label{NS}\ba{c} v_t+v\cdot\n v+\n\pi=R^{-1}\Delta v\,,\;\n \cdot v=0\,,\mbox{ in }(0,T)\times\OO\,,\VS v=0\,,\mbox{ on }(0,T)\times\po\,,\;v=v_0\mbox{ on }\{0\}\times\OO\,.\ea\ee We give the definition of weak solution to problems \rf{NSD} and \rf{NS}. It is immediate to understand that it differs in the problems by a coefficient, that is $\nu$ or $\frac1R$. We opt for the latter, being the latter carried on in the following. 
  \par \begin{defi}\label{WS}{\sl A field $v:(0,\infty)\times \OO\to \R^3$ is said a weak solution to problem \rf{NS} if, for all $T>0$, we have
\begin{itemize}\item [\rm i.] $ v\in L^{\infty}(0,T;J^2(\OO))\cap L^2(0,T;J^{1,2}(\OO))$\,.\item[\rm ii.] $v$ satisfies the integral equation \be\label{IE} (v(t),\varphi(t))=(v_0,\varphi(0))+\intll0t\Big[(v,\varphi_\tau)-R^{-1}(\n v,\n\varphi)+(v\cdot\n\varphi,v)]d\tau\,,\ee for all $\varphi\in C_0^\infty([0,T)\times\OO)$ with $\n\cdot\varphi=0$\,.
\item[\rm iii.] $(v(t),\varphi)$ is a continuous function on $(0,T)$, and $\displ\lim_{t\to0}(v(t),\varphi)=(v_0,\varphi)$\,.
\end{itemize}}\end{defi}
 \par It is well known that for all $v_0\in J^2(\OO)$ there exists a weak solution to problem \rf{NSD}, such a solution enjoys a partial regularity that is expressed by means of the {\it th\'{e}or\`{e}me de structure}. 
Hence, the following result, initially  stated for the Navier-Stokes Cauchy problem associated to \rf{NSD}, is proved
\begin{tho}\label{TS}{\hskip-0.1cm\rm\bf[Leray]}\,-\,{\sl Let $v_0\in J^2(\OO)$. Then there exists a weak solution $v(t, x)$ to   problem \rf{NSD} such that \be\label{TS-I}\ba{c}\displ\dm v(t)\dm_2^2+2\nu\intll st\dm \n v(\tau)\dm_2^2d\tau\leq \dm v(s)\dm_2^2,\mbox{ for all }t>s\,,\mbox{ a. e. in } s>0 \mbox{ and for }s=0\,,\VS v\in C(\theta_\ell,T_\ell;J^{1,2}(\OO))\cap L^2(\theta_\ell,T_\ell;W^{2,2}(\OO))\,,\;v_t\in L^2(\theta_\ell,T_\ell;L^2(\OO)) \,,\VS v\in C(\theta,\infty;J^{1,2}(\OO))\cap L^2(\theta,\infty;W^{2,2}(\OO))\,,\;v_t\in L^2(\theta,\infty;L^2(\OO))\,,\ea\ee where $\theta\leq c \nu^{-5}\dm v_0\dm_2^4$, $(\theta,\infty)\cap(\theta_\ell,T_\ell)=\emptyset$ for all $\ell$,  $(\theta_\ell,T_\ell)\cap(\theta_{\ell'},T_{\ell'})=\emptyset$ for all $\ell\ne\ell'$, the set $\R^+-\{{\underset {\ell\in\N}\cup}(\theta_\ell,T_\ell)\cup(\theta,\infty)\}$ has zero Lebesgue measure\,.} \end{tho}\bp The result in the case of the Cauchy problem can be found in \cite{Lr}. The case of the initial boundary value problem was proved by several authors. For the Cauchy problem, for the IBVP in bounded or in exterior domains the bound for $\theta$ is    the same.\ep We stress that in Theorem\,\ref{TS} the upper bound of $\theta$ has the dimension of the time.
\par We stress that in \cite{Sc} it is proved that $\mathcal H^{\frac12}(\R^+-\{{\underset {\ell\in\N}\cup}(\theta_\ell,T_\ell)\cup(\theta,\infty)\})=0$\,, where $\mathcal H^{\frac12}$ denotes the Hausdorff measure of exponent $\frac12$.
\par Recently, some authors look for an improvement of the original result by Leray. We limit ourselves to quote the papers \cite{FG} and \cite{MrPl-II}. Their results are different for the goals. \par In the paper \cite{FG}, in the special case of  $\Omega\equiv (0,\pi)^3$, with space periodic conditions,   the authors in particular prove a new upper bound for  $\theta $, that is $\theta\leq c\dm v_0\dm_2^2\,.$ \par In the paper \cite{MrPl-II}, the authors furnish a new proof of the {\it th\'{e}or\`{e}me de structure}, based on the regularity of the   sequence approximating the weak solution and the uniqueness of the weak limit. An advantage of this new approach is the fact that the result can be extended in other context of parabolic problems where the Leray approach does not work, see in this connection the paper  \cite{MrPl-III}. \par We set $$\alpha:=2R^3C_S^4c^4\,,\mbox{\; \;}\beta:=RC_S^2c^2\,,\mbox{ \;and\; }\gamma:=C_P^{-1}\,,$$ where $c$ is the constant that appears in \rf{E-I} and $C_S$ is the  constant related to Sobolev's inequality $\dm g\dm_6\leq C_S\dm \n g\dm_2$, and $C_P$ is the constant related to the Poincar\'{e} inequality\footnote{\,Since we are considering  dimensionless quantities,  the Poincar\'e constant is dimensionless number $C_P$.}.\par Here, we are interested to prove the following result
\begin{tho}\label{CT}{\sl \sl Let $v_0\in J^2(\OO)$. Then there exists a weak solution $v(t, x)$ to problem \rf{NS} such that \be\label{TS-I}\ba{c}\displ\dm v(t)\dm_2^2+\sfrac2R\intll st\dm \n v(\tau)\dm_2^2d\tau\leq \dm v(s)\dm_2^2,\mbox{ for all }t>s\,,\mbox{ a. e. in } s>0 \mbox{ and for }s=0\,,\VS v\in C(\theta_\ell,T_\ell;J^{1,2}(\OO))\cap L^2(\theta_\ell,T_\ell;W^{2,2}(\OO))\,,\;v_t\in L^2(\theta_\ell,T_\ell;L^2(\OO)) \,,\VS v\in C(\theta_0,\infty;J^{1,2}(\OO))\cap L^2(\theta_0,\infty;W^{2,2}(\OO))\,,\;v_t\in L^2(\theta_0,\infty;L^2(\OO))\,,\ea\ee where $(\theta_0,\infty)\cap(\theta_\ell,T_\ell)=\emptyset$ for all $\ell$, $(\theta_\ell,T_\ell)\cap(\theta_{\ell'},T_{\ell'})=\emptyset$ for all $\ell\ne\ell'$, the set $\R^3_+-\{{\underset {\ell\in\N}\cup}(\theta_\ell,T_\ell)\cup(\theta_0,\infty)\}$ has zero Lebesgue measure and $\theta_0\in (s,s+\frac\pi4[(1-\eta)(\alpha+\beta)]^{-1} )$\,, where   the istant $s=0$ holds, provided that $R(\alpha+\beta)\dm v_0\dm_2^2\leq \frac\pi2$, the istant $s:=\frac R{2\gamma^2}\log[\frac{4R}{\pi }(\alpha+\beta)\dm v_0\dm^2_2]>0$ holds otherwise.} \end{tho}
\vskip0.1cm {\bf The New Result -}  {\sl The instants  determined in Theorem\,\ref{CT} are dimensionless quantities. Then, related to $\theta_0$, recalling our scaling, we get:
$$\ba{l}\theta:=\mbox{\rm\texttt{T}}\theta_0\in (\mbox{\rm\texttt{T}}s, \mbox{\rm\texttt{T}}s+\frac\pi4(1-\eta)^{-1}\mbox{\rm\texttt{T}}(\alpha+\beta)^{-1} )\,,\VS\mbox{\rm\texttt{T}}s=\sfrac{\mbox{\small\rm\texttt{L}}}{\mbox{\small\rm\texttt{V}}}\sfrac R{2\gamma^2}\log[\sfrac{4R}{\pi }(\alpha+\beta)\dm v_0\dm^2_2]=\sfrac{{\mbox{\small\rm\texttt{L}}}^2}{2\nu\gamma^2}\log[\sfrac{4R}{\pi }(\alpha+\beta)\dm v_0\dm^2_2]\,,\VS \mbox{\rm\texttt{T}}(\alpha+\beta)^{-1}=\sfrac{\mbox{\small\rm\texttt{L}}}{\mbox{\small\rm\texttt{V}}}(\alpha+\beta)^{-1}=\sfrac{\mbox{\small\rm\texttt{L}}}{\mbox{\small\rm\texttt{V}}R^3C_S^4c^4+\mbox{\small\rm\texttt{V}}RC_S^2c^2}=\sfrac{\mbox{\small\rm\texttt{L}}}{\sfrac{\mbox{\small\rm\texttt{V}}^4\mbox{\small\rm\texttt{L}}^3}{\nu^3}C_S^4c^4+\sfrac{\mbox{\small\rm\texttt{V}}^2\mbox{\small\rm\texttt{L}}}{\nu}C_S^2c^2}\,,\ea$$ these quantities, being the kinetic viscosity of dimension  (\hskip0.05cm${\texttt{\rm length})^2\hskip-0.05cm/\texttt{\rm time}}$, have the dimensions of the time variable. \par The result by Leray and the result of this note furnish a different upper bound for instant $\theta$. \par If we fix the domain $\OO$ and the kinetic viscosity $\nu$, we distinguish a different behavior for the upper bound of $\theta$. In the case of small values of $L^2$-norm of the initial datum (in our dimensionless analysis means small Reynolds number) the result by Leray is better, in the sense that its behavior is faster to zero. Instead, in our case a priori the upper bound become large for small $\dm v_0\dm_2$. \par Conversely,  for large values of $L^2$-norm of the initial datum (in our dimensionless analysis means large Reynolds number) the upper bound given by Leray is faster divergent than the one found in this note. Actually,  for large $R$ the interval of $\theta$ from one side has the endpoint $s$ with logarithmic behavior  and from another side the size of the interval tends to zero, that is $\theta$ like $s$. } \vskip0.1cm
 \par 
The previous estimate for instant $\theta$ is the chief result of the note, and, to the best of our knowledge, the endpoint $\theta$ has  the best upper bound for large data.\par Theorem\,\ref{CT} holds for  $\Omega\equiv (0,\pi)^3$, with space periodic conditions, too. The proof needs no change. Since,   the quoted setting  is very particular both in notations and in some analytic questions,   for the sake of brevity, any refinement to  this special domain is omitted.  \par We restrict our result to the case of $\OO$ bounded for two different reasons. The former is connected to the fact that the estimate is of logarithmic kind in terms of the initial datum in $L^2$-norm. The latter is connected to the fact that    in exterior domains we conjecture the estimate to be in   algebraic form of the norms of the initial datum, that is larger than the logarithmic estimate, in any case better of the Leray's result, but we need to require that the initial datum is in some $L^q$ with $q\in[1,2)$. 
\par This  digression on the estimates makes to understand that the proof is connected with the decay of the kinetic energy. This property holds with no further regularity for  a weak solution. We conclude the remark claiming that  the proof  follows the arguments lines of the one obtained  in  recent paper \cite{MrPl-II}. In such  a way we are able to apply the result in another context too. 
\par The plan of the paper is the following. In Sect.\,2, we recall some results related to some a priori estimates and we develop the Leray approach to the construction of the weak solution. In Sect.\,3 we conclude the proof of Theorem\,\ref{CT}.
\section{\large Preliminary results and the Leray approximating sequence} As already said in Sect.\,1, for the sake of brevity, we assume $\OO$ bounded and smooth. However, the regularity can be suitably relaxed with no change of the proof.\par  We denote by $\mathscr C_0(\OO)$ the set of smooth functions $u$ with compact support in $\OO$ and divergence free. By the symbol $J^2(\OO)$ and $J^{1,2}(\OO)$ we mean the completion of $\mathscr C_0(\OO)$ in $L^2(\OO)$ and in $W^{1,2}(\OO)$, respectively. 
\begin{lemma}\label{EL-I}{\sl Let $u\in W^{2,2}(\OO)\cap J^{1,2}(\OO)$. Then there exists a constant $c$  independent of the size of $\OO$ and $u$ such that\be\label{E-I}\dm D^2u\dm_2\leq c(\dm P\Delta u\dm_2+\dm \n u\dm_2)\,,\mbox{\; and \;}\dm \n u\dm_3\leq c(\dm P\Delta u\dm_2^\frac12\dm \n u\dm_2^\frac12+\dm \n u\dm_2)\,.\ee}\end{lemma}
\bp Estimate \rf{E-I} is typical of domain with non-compact boundary, the estimates are due to Heywood, see Lemma\,1 in  \cite{Hy}. \ep
The following lemma is part of a result proved in \cite{CGM-I} Sect.\,6.3.1.\,. \begin{lemma}\label{LFR}{\sl Let $\{v^m\}$ be bounded in $ L^2(0,T;J^{1,2}(\OO))\cap L^\infty(0,T;J^2(\OO))$ and let $\{D^2v^m\}$ be bounded in $L^q(0,T;L^2(\OO))$, for some $q\in(\frac12,2]$.   Denote by $v$ the weak limit in $L^2(0,T;W^{1,2}(\OO))$, then, almost everywhere in $t>0$, we get
$\displ\lim_m\dm \n v^m(t)-\n v(t)\dm_2=0$ \,.}\end{lemma}
\bp Applying  H\"older's inequality, we get
$$\dm \n v^n(\tau)-\n v^p(\tau)\dm_2^2=-(P\Delta (v^n(\tau)-v^p(\tau)),v^n(\tau)-v^p(\tau))\leq \dm P\Delta (v^n(\tau)-v^p(\tau))\dm_2\dm v^n(\tau)-v^p(\tau)\dm_2 \,.$$ Integrating on $(0,T)$, we arrive at
\be\label{SPC}\ba{ll}\displ\intll0T\dm \n v^n(\tau)-\n v^p(\tau)\dm_2^p\hskip-0.2cm&\displ\leq \intll0T\dm P\Delta v^n(\tau)-P\Delta v^p(\tau)\dm_2^\frac p2\dm v^n(\tau)-v^p(\tau)\dm_2^\frac p2d\tau\\&\displ\leq \Big[\intll0T \dm P\Delta v^n(\tau)-P\Delta v^p(\tau)\dm_2^qd\tau\Big]^\frac p{2q}\Big[\dm v^n(\tau)-v^p(\tau)\dm_2^\frac{pq}{2q-p}d\tau\Big]^\frac{2q-p}{2q}\\&\displ\leq M\Big[\intll0T\dm v^n(\tau)-v^p(\tau)\dm_2^\frac{pq}{2q-p}d\tau\Big]^\frac{2q-p}{2q} \,,\ea\ee provided that $p\in[1,2q)$. By virtue of our assumptions and the Friederich's lemma, \cite{Ld},  the last estimate ensures that  $\{\n v^n\}$ satisfies a Cauchy condition in $L^p(0,T;L^2(\OO))$. Therefore, the strong convergence of $\{\n v^m\}$ to $\n v$  in $L^p(0,T;L^2(\OO))$, that, finally, implies the thesis of the lemma. 
\ep Let us consider the following initial boundary value problem:
\be\label{NSm}\ba{c}v^m_t+\mathbb J^m(v^m)\cdot\n v^m+\pi^m=\ssfrac1R\Delta v^m\,,\quad\n\cdot v^m=0\,,\mbox{\, in\, }(0,T)\times\OO\,,\VS
v^m=0\mbox{\, on\, }(0,T)\times\po\,,\mbox{\, and\, }v^m=v_0\mbox{\, on\, }\{0\}\times\OO\,,\ea\ee
where $\mathbb J_n\equiv J_{\frac1n}$ that is the Friederich mollifier.
\begin{tho}\label{LE}{\sl Let  $v_0\in J^2(\OO)$. Then there exists a weak solution $(v^m,\pi^m)$ to problem \rf{NSm}. Moreover, with a bound not uniform with respect to $m\in\N$ and $\eta>0$, the solution enjoys  the estimates 
\be\label{LE-I}\ba{c}  v^m\in C(\eta,T;J^{1,2}(\OO))\cap L^2(\eta,T;W^{2,2}(\OO))\,,\VS v^m_t,\,\n\pi^m\in L^2(\eta,T;L^2(\OO))\,. \ea\ee Finally, uniformly in $m\in\N$   the following hold
\be\label{EO}(v^m(t),\varphi)\in C(0,T)\,,\mbox{ for all }\varphi\in \mathscr C(\OO)\,,\;\displ\lim_{t\to0}\dm v^m(t)-v_0\dm_2=0\,,\ee
\be\label{EI}\dm v^m(t)\dm_2^2+\sfrac2R\intll st\dm\n v^m(\tau)\dm_2^2d\tau=\dm v^m(s)\dm_2^2\,,\mbox{ for all }t>s\geq0\,,\ee \be\label{ED}\dm v^m(t)\dm_2^2\leq \dm v_0\dm_2^2\exp[-\ssfrac2R\gamma^2 t]\,,\mbox{ for all }t>0\,,\ee
where we set $\gamma:=  C_P^{-1}$. 
}\end{tho}
\bp We follow the classical construction, see \cite{Pr} and \cite{Hy}, by means of the sequence $\{a_\ell\}$ of egeinvectors  to the Stokes operator $P\Delta a_\ell=-\lambda^\ell a_\ell\,,$ $\ell\in\N$.
We consider $\{a_\ell\}$ orthonormal in $J^2(\OO)$. Setting $u^m_p:={\underset{\ell=1}{\overset p\sum}}c^\ell_p(t)a_\ell$, we realize the   system of ordinary differential equations:  \be\label{GA}({v^m_p\hskip-0.1cm}_t,a_\ell)+\ssfrac1R(\n v^m_p,\n a_\ell)=-(\mathbb J_m(v^m_p)\cdot\n v^m_p,a_\ell)\,,\;\ell=1,\cdots,p\,,\ee with $v^m_p(0)=
{\underset{\ell=1}{\overset p\sum}}c_p^\ell(0)a_\ell$ and $c_p^\ell(0):=(v_0,a_\ell)$, for $\ell=1,\cdots p$\,. We can solve globally in time the solutions to system \rf{GA} thanks the energy relation that holds  for all $t>0$ and $p\in\N$:
\be\label{EIP}\dm v_0\dm_2^2\geq \!\dm v_p^m(0)\dm_2\!=\!\mbox{${\underset{\ell=1}{\overset p\sum}}$}\hskip-0.03cm(c^\ell_p(0))^2\!=\!\ssfrac1R\!\!\intll0t\!\!\dm \n v^m_p(\tau)\dm_2^2d\tau+\dm v^m_p(t)\dm_2^2=\!\ssfrac 1R\!\!\intll0t\!\!\dm\n v^m_p(\tau)\dm_2^2d\tau+\!\mbox{${\underset{\ell=1}{\overset p\sum}}$}(c^\ell_p(t))^2.\ee
Multiplying by $R^{-1}\lambda_\ell c^\ell_p(t)+\dot c^\ell_p(t)$, and summing on $\ell=1,\cdots,p$, we get
$$\dm R^{-1}P\Delta v^m_p(t)-{v^m_p\hskip-0.1cm} _t(t)\dm_2^2=(\mathbb J_m(v^m_p)\cdot\n v^m_p,R^{-1}\Delta v^m_p(t)+{v^m_p\hskip-0.1cm}_t(t))\,,$$
that easily furnishes 
$$ \dm R^{-1}P\Delta v^m_p(t)-{v^m_p\hskip-0.1cm} _t(t)\dm_2^2\leq\dm\mathbb J_m(v^m_p)\cdot\n v^m_p\dm_2^2\,.$$
Hence, we arrive at
\be\label{GRE}\sfrac d{dt}\dm\n v^m_p(t)\dm_2^2+\ssfrac{1\hskip-0.1cm}{R}\dm P\Delta v^m_p(t)\dm_2^2+R\dm {v^m_p\hskip-0.1cm} _t(t)\dm_2^2\leq c(m)R\dm v^m_p(t)\dm_6^2\dm \n v^m_p(t)\dm_2^2\leq c(m)R\dm\n v^m_p(t)\dm_2^4\,.\ee This last integrated on $(s, t)$, via the energy estimate, furnishes
$$\dm \n v^m_p(t)\dm_2^2\leq \dm \n v^m_p(s)\dm_2^2\exp[c(m)\dm v_0\dm_2^2]$$
Integrating on $(0,t)$, again via energy estimate, we also deduce
\be\label{GRE-I}t\dm \n v^m_p(t)\dm_2^2\leq \ssfrac R2\dm v_0\dm_2^2\exp[c(m)\dm v_0\dm_2^2]\,.\ee
Considering again  estimate \rf{GRE}, employing on the right-hand side \rf{GRE-I}, an integration on $(\eta,T)$ furnishes uniformly in $p\in\N$ and $t>\eta$, for all $\eta>0$, the estimate
\be\label{DSP}\intll\eta t\Big[R^{-1}\dm P\Delta v^m_p(\tau)\dm_2^2+R\dm {v^m_p\hskip-0.1cm} _t(\tau)\dm_2^2\Big]d\tau\leq \ssfrac{R^2\hskip-0.1cm}2Rc(m)\eta^{-1}\dm v^m_0\dm_2^4\exp[c(m)\dm v_0\dm_2^2\,.\ee This last leads to the existence of $v^m$ enjoying estimate \rf{LE-I}, for all $m\in\N$, with a constant depending on $m\in\N$. Now, we are going to prove the estimates \rf{EO}-\rf{ED}. The continuity expressed in \rf{EO} is classical in the Navier-Stokes mathematical theory, see {\it e.g.} \cite{Ld}. In the case of \rf{EI}, we start from the energy relation \rf{EIP}. By  difference we get 
$$\dm v^m_p(t)\dm_2^2+\ssfrac2R\intll st\dm \n v^m_p(\tau)\dm_2^2d\tau=\dm v^m_p(s)\dm\,.$$ Taking  estimates \rf{GRE-I}-\rf{DSP}  into account, via \rf{SPC} written with $q=2$, employing the Rellich-Kolmogorov theorem, letting $p\to\infty$, on interval $(s,t)$ with $s>0$, we arrive at
$$\dm v^m(t)\dm_2^2+\ssfrac2R\intll st\dm \n v^m(\tau)\dm_2^2d\tau=\dm v^m(s)\dm_2^2\,.$$ On the other hand, the continuity in norm for $t=0$ expressed
in \rf{EO}, allows us to consider the energy relation achieved for all $t>s\geq0$. This proves \rf{EI}. Estimate \rf{ED} is classical, see {\it e. g.} \cite{St}. The theorem is completely proved.\ep 
The following lemma has proved in \cite{CGM-I}, for the sake of completeness we furnish the proof.
 \begin{lemma}\label{CGM}{\sl   Let $\{(v^m,\pi^m)\}$ the sequence furnished in Theorem\,\ref{LE}, then we get
\be\label{CGM-I}\intll0T\dm D^2v^m(t)\dm_2^\frac23dt\leq c(\dm v_0\dm_2^2,T)\,,\mbox{ for all }m\in\N\,.\ee  
}\end{lemma}
\bp From system \rf{NSm} we deduce the identity in norm:
$$\dm v^m_t(t)-R^{-1}P\Delta v^m(t)\dm_2^2=\dm P\mathbb J_m(v^m(t))\cdot\n v^m(t)\dm_2^2$$ Developing the $L^2$-norm on  the left-hand side, and applying the Sobolev inequality for the right-hand side, we arrive at
$$\ssfrac1R\sfrac d{dt}\dm \n v^m(t)\dm_2^2+\ssfrac{1 }{R^2\hskip-0.1cm}\hskip0.1cm\dm P\Delta v^m(t)\dm_2^2+\dm v^m_t(t)\dm_2^2\leq \dm \mathbb J_m(v^m(t))\cdot\n v^m(t)\dm_2^2\leq \dm v^m(t)\dm_6^2\dm \n v^m(t)\dm_3^2\,.$$ 
Employing estimate \rf{E-I}, we get
$$\ba{ll}\sfrac d{dt}\dm \n v^m(t)\dm_2^2+\!\ssfrac1R\dm P\Delta v^m(t)\dm_2^2+\!R\dm v^m_t(t)\dm_2^2\hskip-0.25cm&\leq 2RC_S^2c^2\dm \n v^m(t)\dm_2^3 \dm P\Delta v^m(t)\dm_2\!+2RC_S^2c^2\dm \n v^m(t)\dm_2^4\VSE\hskip-1cm
\leq 2R^3C^4_Sc^4\dm\n v^m(t)\dm_2^6+\!RC_S^2c^2\dm \n v^m(t)\dm_2^4\!+\!\ssfrac1{2R}\dm P\Delta v^m(t)\dm_2^2\,,\ea $$
Multiplying by $(1+\dm\n v^m(t)\dm_2)^{-2}$ we arrive at
$$(1+\dm\n v^m(t)\dm_2^2)^{-2}\sfrac d{dt}\dm \n v^m(t)\dm_2^2+\sfrac12\sfrac{\dm P\Delta v^m(t)\dm_2^2}{(1+\dm \n v^m(t)\dm_2^2)^2}\leq  c\dm \n v^m(t)\dm_2^2\,.$$ Integrating on $(s,T)$ and applying H\"older's inverse inequality with exponent $-\frac12,\,\frac13$, we obtain the estimate
$$\ba{ll}\displ\Big[\intll sT\dm P\Delta v^m(\tau)\dm_2^\frac23d\tau\Big]^3\hskip-0.3cm&\displ\leq \Big[(1+\dm\n v^m(t)\dm_2^2)^{-1}+c\dm v_0\dm_2^2\Big]\Big[\intll sT(1+\dm\n v^m(\tau)\dm_2^2)d\tau\Big]^2\VSE\leq 
\Big[1+c\dm v_0\dm_2^2\Big]\Big[\intll 0T(1+\dm\n v^m(\tau)\dm_2^2)d\tau\Big]^2\,,\ea$$ that, letting $s\to0$, leads to the thesis.
\ep
The following lemma is crucial for our aims.
\begin{lemma}\label{CL}{\sl Let $t_0\geq0$,  $v(t_0)\in J^2(\OO)$ with $R(\alpha+\beta)\dm v(t_0)\dm_2^2\leq\frac\pi2$ and assume $\eta\in (0,1)$. Then there exists $\theta_0\in (t_0,t_0+\frac\pi4[R(1-\eta)(\alpha+\beta)]^{-1} )$ such that the weak solutions $(v^m,\pi^m)$ furnished in Theorem\,\ref{LE}, uniformly in $m\in\N$,  enjoys the estimates:
\be\label{TH-I}\ba{c}  v^m\in C(\theta_0,T;J^{1,2}(\OO))\cap L^2(\theta_0,T;W^{2,2}(\OO))\,,\mbox{ for all }T>\theta_0\,,\VS v^m_t,\,\n\pi^m\in L^2(\theta_0,T;L^2(\OO))\,,\mbox{ for all }T>\theta_0\,. \ea\ee Finally, uniformly in $m\in\N$,   the following hold
\be\label{TH-II}(v^m(t),\varphi)\in C(t_0,T)\,,\mbox{ for all }\varphi\in \mathscr C(\OO)\mbox{ and }T>0\,,\;\displ\lim_{t\to t_0}\dm v^m(t)-v(t_0)\dm_2=0\,,\ee
\be\label{TH-III}\dm v^m(t)\dm_2^2+\sfrac2R\intll st\dm\n v^m(\tau)\dm_2^2d\tau=\dm v^m(s)\dm_2^2\,,\mbox{ for all }t>s\geq t_0\,,\ee \be\label{TH-IV}\dm v^m(t)\dm_2^2\leq \dm v_0\dm_2^2\exp[-\ssfrac2R\gamma^2 t]\,,\mbox{ for all }t>t_0\,.\ee  
}\end{lemma}
\bp By virtue of Theorem\,\ref{LE} the solutions to problem \rf{NSm} are defined for all $t>t_0$. Now, we look for estimates that are independent of $m\in\N$. We claim that given a $\eta\in(0,1)$,  for all $m\in\N$, there exists a $t_m\in (t_0,t_0+\frac {R\dm v(t_0)\dm_2^2}{2(1-\eta)} )$ such that $\dm \n v^m(t_m)\dm_2^2<1-\eta\,.$ By contradiction, we assume that $$\dm \n v^m(t)\dm_2^2\geq1-\eta\,,\mbox{ for all }t\in(t_0,t_0+\ssfrac {R\dm v(t_0)\dm_2^2}{2(1-\eta)})\,.$$ Then, we also get
$$\dm v(t_0)\dm_2^2=\sfrac2R(1-\eta)\hskip-0.5cm\intll {t_0}{t_0+\frac {R\dm v(t_0)\dm_2^2}{2(1-\eta)} }\hskip-0.5cmd\tau\leq\sfrac2R\hskip-0.5cm\intll {t_0}{t_0+\frac {R\dm v(t_0)\dm_2^2}{2(1-\eta)} }\hskip-0.5cm\dm\n v^m(\tau)\dm_2^2d\tau\,,$$
that contradicts the validity of \rf{EI}  for all $m\in\N$ and $s=t_0$. From system \rf{NSm} we deduce the identity in norm:
$$\dm v^m_t(t)-R^{-1}P\Delta v^m(t)\dm_2^2=\dm P\mathbb J_m(v^m(t))\cdot\n v^m(t)\dm_2^2$$ Developing the $L^2$-norm on  the left-hand side, and applying the Sobolev inequality for the right-hand side, we arrive at
$$\sfrac d{dt}\dm \n v^m(t)\dm_2^2+R^{-1}\dm P\Delta v^m(t)\dm_2^2+R\dm v^m_t(t)\dm_2^2\leq R \dm \mathbb J_m(v^m(t))\cdot\n v^m(t)\dm_2^2\leq R\dm v^m(t)\dm_6^2\dm \n v^m(t)\dm_3^2\,.$$ 
Employing estimate \rf{E-I}, we get
$$\ba{ll}\sfrac d{dt}\dm \n v^m(t)\dm_2^2+\!\ssfrac1R\dm P\Delta v^m(t)\dm_2^2\!+R\dm v^m_t(t)\dm_2^2\hskip-0.3cm&\leq 2RC_S^2c^2\dm \n v^m(t)\dm_2^3\dm P\Delta v^m(t)\dm_2\!+2RC_S^2c^2\dm \n v^m(t)\dm_2^4\VSE\hskip-1cm
\leq 2RC^2_Sc^2\!\Big[R^2C^2_Sc^2\dm\n v^m(t)\dm_2^6+\!\dm \n v^m(t)\dm_2^4\Big]\!+\!\ssfrac1{2R}\dm P\Delta v^m(t)\dm_2^2.\ea $$
From the last differential inequality, with obvious meaning of constant $\alpha$ we deduce that
\be\label{NDR} \sfrac d{dt}\dm \n v^m(t)\dm_2^2+\sfrac1{2R}\dm P\Delta v^m(t)\dm_2^2+R\dm v^m_t(t)\dm_2^2\leq \alpha\dm\n v^m(t)\dm_2^6+2RC_S^2c^2\dm \n v^m(t)\dm_2^4\,,\ee and then the following
$$(1+\dm\n v^m(t)\dm_2^4)^{-1} \sfrac d{dt}\dm \n v^m(t)\dm_2^2\leq \alpha\dm\n v^m(t)\dm_2^2+\beta\dm \n v^m(t)\dm_2^2\,,$$
where for the last term we toke the inequality $\frac x{1+x^2}\leq \frac12$ on $(0,\infty)$ into account. We denote by $\sigma\in (0,\frac\pi4)$ an angle such that $\tan(\frac\pi4- \sigma)=1-\eta$. Integrating on $(t_m,t)$, being $\dm v^m(t_m)\dm_2^2\leq \dm v(t_0)\dm_2^2$, we get
$$\arctan \dm\n v^m(t)\dm_2^2\!\leq\! \arctan \dm \n v^m(t_m)\dm_2^2+(\alpha\!+\!\beta)\!\intll{t_m}t\!\!\dm\n v^m(\tau)\dm_2^2d\tau\!<\! \frac\pi4-\sigma+\sfrac R2(\alpha\!+\!\beta)\dm v^m(t_0)\dm_2^2\leq \sfrac\pi2-\sigma\,. $$ Hence, the right hand side is bounded by $\frac\pi2-\sigma$ uniformly in $m\in N$ and $t>t_m$. Moreover, being $\dm v(t_0)\dm_2^2\leq \frac\pi2[R(\alpha+\beta)]^{-1}$, we get  $t_m\in(t_0,t_0+\frac\pi4[R(1-\eta)(\alpha+\beta)]^{-1} )$. Hence, the sequence $\{t_m\}$ admits $\displ\sup_mt_m$, say $\theta_0$. Therefore the last estimate holds uniformly  in $m\in\N$ and in $t>\theta_0$, in particular, it implies \be\label{SGU}\dm \n v^m(t)\dm_2^2\leq\tan  (\frac\pi2-\sigma)=:A<\infty\,,\mbox{ for all }m\in\N\mbox{ and }t\geq\theta_0.\ee 
Considering again  estimate \rf{NDR}, then we realize 
$$\dm \n v^m(t)\dm_2^2+\!\intll{\theta_0}t\!\Big[\sfrac12\dm P\Delta v^m(\tau)\dm_2^2+\dm v^m_\tau(\tau)\dm_2^2\Big]d\tau\leq A+\Big[\alpha A^2+\beta A\Big]\dm v(t_0)\dm_2^2\,,\mbox{ for all }m\in\N\mbox{ and }t\geq\theta_0.$$ The lemma is proved.
\ep
The following lemma is well known in literature. Thus, we omit the proof.
\begin{lemma}\label{ERSL}{\sl Let $v(t_0)\in J^{1,2}(\OO)$. Then there exists a unique solution to problem \rf{NS} such that
\be\label{ERS-I}v\in C([t_0,t_0+T;J^{1,2}(\OO))\cap L^2(t_0,t_0+T;W^{2,2}(\OO))\,,v_t,\n\pi\in L^2(t_0,t_0+T;L^2(\OO))\,,\ee
where $(t_0,t_0+T)$ is a maximal interval of existence with $T\geq c\dm \n v(t_0)\dm_2^{-4}$\,.}\end{lemma} 
\section{\large Proof of Theorem\,\ref{CT}}
Let $v_0\in J^2(\OO)$. The energy estimate \rf{EI} for $t>0$ and $s=0$ ensures that the sequence $\{v^m\}$ adimits a weak limit $v$ in $L^2(0,T;J^{1,2}(\OO))$ which is a weak solution to problem \rf{NS}. It is well known that the limit satisfies the energy inequality on strong form \rf{TS-I}$_1$, and the initial datum is assumed  strongly in $L^2$-norm, that is $\displ\lim_{t\to0}\dm v(t)-v_0\dm_2^2=0$. Since this result is classical, we omit any detail. \par Now, we are going to show the {\it th\`eor\'eme de structure}. We follow the idea employed in \cite{MrPl-II}. That is,    $v^m$, almost everywhere in $t>0$, admits an interval  $(t,t+T(t))$, independent of $m\in\N$, such that
\be\label{SFR}v^m\in C(t,t+T(t);J^{1,2}(\OO))\cap L^2(t,t+T(t);W^{2,2}(\OO))\, \;v^m_t\in L^2(t,t+T(t);L^2(\OO))\,.\ee
Denoted by $\ov v$ the limit of the sequence $\{v^m\}$ with respect the metric \rf{SFR}, for the uniqueness of the weak limit $v$, the regularity detected by \rf{SFR} holds for $v$ too.\par We start proving the existence of the interval $(\theta_0,\infty)$. For this goal in the following we distinguish two cases. \par If $\dm v_0\dm_2^2R(\alpha+\beta)\leq\frac\pi2$, estimate \rf{SFR} is already stated in the previous lemma, but   now we have $t_0=0$, $\theta_0\in(0,\frac\pi4[R(1-\eta)(\alpha+\beta)]^{-1})$ and   interval of regularity $(\theta_0,\infty)$. \par So for estimate \rf{SFR} we are going to consider the case of $\dm v_0\dm_2^2R(\alpha+\beta)>\frac\pi2$. Let $\{(v^m,\pi^m)\}$ the sequence of solutions stated in Theorem\,\ref{LE}. We consider estimate \rf{ED}. We fix the instant $s$ such that $e^{-2\frac{\gamma^2\hskip-0.15cm}R s}R(\alpha+\beta)\dm v_0\dm_2^2=\frac\pi2$. Hence, we get   \be\label{VS}\dm v^m(s)\dm_2^2R(\alpha+\beta)\leq\frac\pi2\mbox{ and } s:=\sfrac R{2\gamma^2}\log\big[ \sfrac{2R}\pi {\dm v_0\dm_2^2}{(\alpha+\beta)}\big]\,.\ee Now, for a given $\eta\in(0,1)$  we  prove the existence of an instant $t_m\in[s,s+\frac{R\dm v^m(s)\dm_2^2}{2(1-\eta)}]$ such that $\dm\n v^m(t_m)\dm_2^2< 1-\eta$. The proof is the same given in the previous lemma. That is, $$\dm v^m(s)\dm_2^2=(1-\eta)\sfrac 2R\hskip-0.5cm\intll{s}{s+\frac {R\dm v(s)\dm_2^2}{2(1-\eta)} }\hskip-0.5cmd\tau\leq  \sfrac2R\hskip-0.5cm\intll{s}{s+\frac {R\dm v(s)\dm_2^2}{2(1-\eta)} }\hskip-0.5cm\dm \n v^m(\tau)\dm_2^2dd\tau\,,$$ that contradicts the energy relation \rf{EI}. Hence, there exists $t_m\in[s,{s+\frac {R\dm v(s)\dm_2^2}{2(1-\eta)} }]$ such that $\dm\n v^m(t_m)\dm_2^2< 1-\eta$. Since estimate \rf{ED} implies that $\dm v^m(s)\dm_2^2\!\leq \!e^{-\frac2{R}\gamma^2s}\dm v_0\dm_2^2$, substituting the value of $s$ given in \rf{VS}, we get $\{t_m\}\subset [s,s+\frac\pi4[(1-\eta)(\alpha+\beta)]^{-1}]$. Now, we procede as in the previous lemma. Hence, there exists $\theta_0\in[s,s+\frac\pi4[(1-\eta)(\alpha+\beta)]^{-1} ] $ such that \be\label{SFDS}\ba{l}\displ\dm \n v^m(t)\dm_2^2+\intll{\theta_0}t\Big[\ssfrac12\dm P\Delta v^m(\tau)\dm_2^2+\dm v^m_\tau(\tau)\dm_2^2\Big]d\tau\leq A+\Big[\alpha A^2+\beta A\Big]\dm v(t_0)\dm_2^2\,,\\\hskip10cm\mbox{ for all }m\in\N\mbox{ and }t\geq\theta_0\,, \ea\ee where $A=\tan (\sfrac\pi2-\sigma) > \dm \n v^m(t)\dm_2^2$, for all $t\in(\theta_0,\infty)$.
In both the cases we get \rf{SFDS} that ensure that the sequence $\{v^m\}$ admits a limit $\ov v$  with respect the metrics detected by \rf{SFDS}. For the uniqueness of the weak limit $v$ in $L^2(0,T;J^{1,2}(\OO))$, we realize that $v\equiv\ov v$ on $(\theta_0,\infty)$.\par
Now, the proof considers the cases of $t\in(0,\theta_0)$, As already said we follows the argument lines of paper \cite{MrPl-II}, that we reproduce for the sake of completeness. 
\par Lemma\,\ref{CGM} allows us to apply Lemma\,\ref{LFR} assuming $q=\frac23$. Hence, there exists the set $\mathscr T\subseteq(0,\theta_0)$ such that, for all $t\in\mathscr T$, $\n v^m(t)\to \n v(t)$ strongly in $L^2(\OO)$.
\par Let us consider $t_\alpha\in [0,\theta_0]\cap \mathscr T$. Let $m(t_\alpha)$ be such that $\dm\n v^m(t_\alpha)\dm_2^2\leq \dm \n v(t_\alpha)\dm_2^2+1$.  By virtue of \rf{ERS-I} we can consider an extract $\{v^m\}, m\geq m(t_\alpha)$, that admits a limit $(w,\pi_w)$ regular solution on some maximal interval of the kind $I_\alpha:=[t_\alpha,t_\alpha+T(t_\alpha))\equiv {\underset{m\geq m(t)}\cap}[t_\alpha,t_\alpha +T_m(t_\alpha))$, non empty set since it contains strictly the interval $[t_\alpha,t_\alpha+c(\dm\n v(t_\alpha)\dm_2^2+1)^{-2}]$, being $\dm \n v^m(t_\alpha)\dm_2^{-2}\geq (\dm \n v(t_\alpha)\dm_2^{2}+1)^{-1}$ for all $m\geq m(t_\alpha)$.   By virtue of the uniqueness of the weak limit in  $L^2(0,\theta_0;J^{1,2}(\OO))$, we  get that $v\equiv w$ again. Hence, we have proved a local in time property of regularity for $v$. Let us consider $t_\beta\in [0,\theta_0]\cap \mathscr T-I_\alpha$. By the same previous argument lines,  we can state the existence of $I_\beta$ like maximal  interval of regularity for the weak solution $v$. Now, there are two possibility:
\begin{itemize}\item[i.] $I_\alpha\cap I_\beta=\emptyset$;\item[ii.] $I_\alpha\cap I_\beta\ne\emptyset$, then $I_\beta\supset I_\alpha$.\end{itemize}
We justify the second item. If  $I_\alpha\cap I_\beta\ne\emptyset$, since $t_\beta\notin I_\alpha$, then $t_\beta<t_\alpha$. So that we have $t_\alpha\in I_\beta$. Hence $I_\alpha $ represents an extension of $I_\beta$, that is $t_\beta+T(t_\beta)=t_\alpha+T(t_\alpha)$. In the case of item i. We start again with a new $t_\gamma\in [0,\theta_0]\cap \mathscr T-\{I_\alpha\cup I_\beta\}$. In the case of item ii. we replace $I_\alpha$ with $I_\beta$ and start with $t_\gamma\in [0,\theta_0]\cap \mathscr T-I_\beta$. Iterating the procedure, we state a family of intervals $\{I_\alpha\}_{\alpha\in\mathscr A}$, with $\mathscr A$ set of indexes. We have that $[0,\theta_0]\cap \mathscr T\subseteq {\underset{\alpha\in \mathscr A}\cup}I_\alpha$, with ${\overset\circ{I_\alpha}}\ne\emptyset$ and ${\overset\circ{I_\alpha}}\cap{\overset\circ{ I_\beta}}=\emptyset$ for $\alpha\ne\beta$. The last property ensures that the set of indexes $\mathscr A$ has at most the cardinality of $\N$. The theorem is completely proved.


\begin{thebibliography}{20}
\baselineskip=9pt
 \bibitem{FG}G. Arioli, A. Falocchi, F. Gazzola, {\it On the epochs of irregularity of Leray-Hopf solutions to Navier-Stokes equations}, 
J. of Diff. Eq. {\bf 453} (2026) https://doi.org/10.1016/j.jde.2025.113887
\bibitem{CGM-I}F. Crispo, C.R. Grisanti and P. Maremonti, {\it Some new properties of a suitable weak solution to the Navier-Stokes equations}, in Waves in Flows: The 2018 Prague-Sum Workshop Lectures, series: Lecture Notes in Mathematical Fluids Mechanics, editors: G.P.Galdi, T. Bodnar, S. Necasova, Birkhauser.
\bibitem{Hy}J.G. Heywood, {\it The Navier-Stokes equations: on the existence, regularity and decay of solutions}, Indiana Univ. Math. J. {\bf 29} (1980), no. 5, 639-681. 
\bibitem{Ld}O.A. Ladyzhenskaya, {\it The mathematical theory of viscous incompressible flow,}  Gordon and Breach Sc. Publ., New York-London-Paris, 1969.  
\bibitem{Lr}J. Leray, {\it Sur le mouvement d'un liquide visqueux emplissant l'espace},  Acta Math. {\bf 63} (1934), no. 1, 193-248. 
\bibitem{MrPl-II} P. Maremonti and F. Palma, {\it A new proof of the Th\`eor\'eme de Structure related to a weak solution to the Navier-Stokes equations,} to appear on JMFM, arXiv:2512.05598
\bibitem{MrPl-III} P. Maremonti and F. Palma, {\it On the interaction between a rigid-body and a viscous-fluid: existence of a weak solution and a suitable Th\`{e}or\'eme de Structure}, 
arXiv:2602.11787 
\bibitem{Pr}G. Prodi, {\it Teoremi di tipo locale per il sistema di Navier-Stokes e stabilit\`{a} delle soluzioni stazionarie} (Italian) Rend. Sem. Mat. Univ. Padova {\bf 32} (1962) 374-397. 
\bibitem{St}D. H. Sattinger, {\it The mathematical problem of hydrodynamic stability,} Journal of Mathematics and Mechanics, {\bf 19} (1970) 797-817.
\bibitem{Sc}V. Scheffer, {\it
Turbulence and Hausdorff dimension. Turbulence and Navier-Stokes equations},  Lecture Notes in Math., Vol. {\bf 565} Springer, Berlin-New York, 1976.
\end{thebibliography}
\end{document}